\documentclass[letterpaper]{article} 
\usepackage{aaai23}  
\usepackage{times}  
\usepackage{helvet}  
\usepackage{courier}  
\usepackage[hyphens]{url}  
\usepackage{graphicx} 
\usepackage{natbib}  
\usepackage{caption} 
\usepackage{algorithm}
\usepackage{algorithmic}

\usepackage{newfloat}
\usepackage{listings}
\DeclareCaptionStyle{ruled}{labelfont=normalfont,labelsep=colon,strut=off} 
\floatstyle{ruled}
\newfloat{listing}{tb}{lst}{}
\floatname{listing}{Listing}
\title{IdeaReader: A Machine Reading System for Understanding the Idea Flow of Scientific Publications}
\author {
    Qi Li\textsuperscript{\rm 1},
    Yuyang Ren\textsuperscript{\rm 1}, 
    Xingli Wang\textsuperscript{\rm 1}, 
    Luoyi Fu\textsuperscript{\rm 1}, 
    Jiaxin Ding\textsuperscript{\rm 1}, 
    Xinde Cao\textsuperscript{\rm 1}, 
    Xinbing Wang\textsuperscript{\rm 1\footnotemark[2]}, 
    Chenghu Zhou\textsuperscript{\rm 2}
}
\affiliations {
    \textsuperscript{\rm 1} Shanghai Jiao Tong University\\
    \textsuperscript{\rm 2} Institute of Geographic Sciences and Natural Resources Research, Chinese Academy of Sciences\\
    \{liqilcn,renyuyang,xingliwang,yiluofu,jiaxinding\}@sjtu.edu.cn, xdcaosjtu@outlook.com, xwang8@sjtu.edu.cn, zhouch@lreis.ac.cn
}

\usepackage{bibentry}

\begin{document}
\nocopyright  

\maketitle
\renewcommand{\thefootnote}{\fnsymbol{footnote}} 
\footnotetext[2]{Corresponding authors.}

\begin{abstract}
Understanding the origin and influence of the publication's idea is critical to conducting scientific research. However, the proliferation of scientific publications makes it difficult for researchers to sort out the evolution of all relevant literature. To this end, we present IdeaReader, a machine reading system that finds out which papers are most likely to inspire or be influenced by the target publication and summarizes the ideas of these papers in natural language. Specifically, IdeaReader first clusters the references and citations (first-order or higher-order) of the target publication, and the obtained clusters are regarded as the topics that inspire or are influenced by the target publication. It then picks out the important papers from each cluster to extract the skeleton of the idea flow. Finally, IdeaReader automatically generates a literature review of the important papers in each topic. Our system can help researchers gain insight into how scientific ideas flow from the target publication's references to citations by the automatically generated survey and the visualization of idea flow. Code is available at \url{https://github.com/liqilcn/IdeaReader}.
\end{abstract}

\section{Introduction}

An excellent scientific paper usually serves as a connecting link in the development of science. It is usually first inspired by previous work, published after incorporating the authors' idea and then influences subsequent work. Understanding the origin and influence of the publication's idea is an indispensable step in conducting scientific research. For novice or cross-disciplinary researchers, they need to sort out the evolution of the massive literature's idea to enter a new field \cite{zha2019mining}. When writing a paper, researchers also need to figure out the evolution of the related work's idea to show the advantages of their work in the article \cite{hoang2010towards, hu2014automatic}. However, the proliferation of scientific publications has made it increasingly difficult for scholars to understand the contextual research of all relevant papers \cite{chu2021slowed}. 

In this paper, we propose IdeaReader, a machine reading system that helps researchers quickly understand the evolution of scientific publications' ideas. Our system (available at \url{https://agkg.acemap.cn}) has the following three characteristics: (1) IdeaReader can find out which articles are most likely to inspire or be influenced by the target publication from the references and citations of the target publication (first-order or higher-order). (2) IdeaReader can utilize natural language to automatically summarize the ideas of the target publication's contextual research. (3)  IdeaReader can visualize the flow of the ideas in important literature associated with the target publication.

\section{Related Work}

Our work is mainly related to the construction of scientific evolution roadmaps and automatic related work generation. \textbf{Construction of scientific evolution roadmaps:} \citeauthor{he2009detecting} (2009) adapt the LDA model to the citation network to construct topic evolution roadmaps of scientific literature. \citeauthor{zha2019mining} (2019) use the tabular data in the paper as weakly supervised signals to extract algorithmic roadmaps from massive publications. \citeauthor{yin2021mrt} (2021) utilize spectral clustering to build paper evolution roadmaps from the query paper's reference. \textbf{Automatic related work generation:} \citeauthor{hoang2010towards} (2010) use two separate strategies to generate general and detailed sentences of the related work text. \citeauthor{hu2014automatic} (2014) propose an optimization method to select important sentences from both the target paper and the reference papers to generate a related work section for a target paper. \citeauthor{chen2021capturing} (2021) design a abstractive related work generation model by automatically capturing the content dependency between multiple input documents during training. 

However, the scientific evolution roadmaps can not reveal the specific research content of the corresponding papers. Our system utilizes the extracted skeleton of idea flow to guide survey generation, which can directly present the ideas of the target paper's contextual research.

\section{System Description}

\begin{figure*}[t]
\centering
\includegraphics[width=1\textwidth]{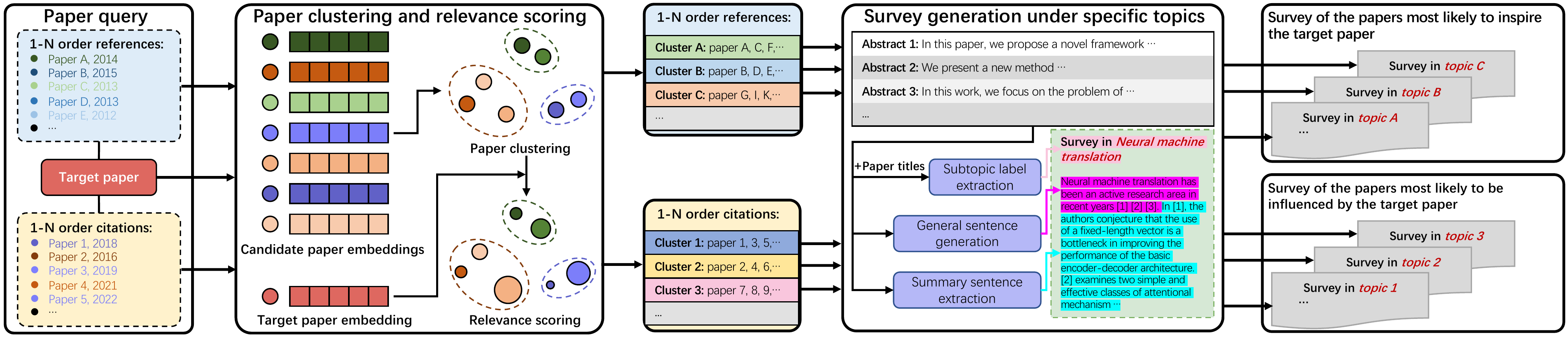} 
\caption{The algorithm pipeline of the IdeaReader.}
\label{fig1}
\end{figure*}
\begin{figure}[t]
\centering
\includegraphics[width=1\columnwidth]{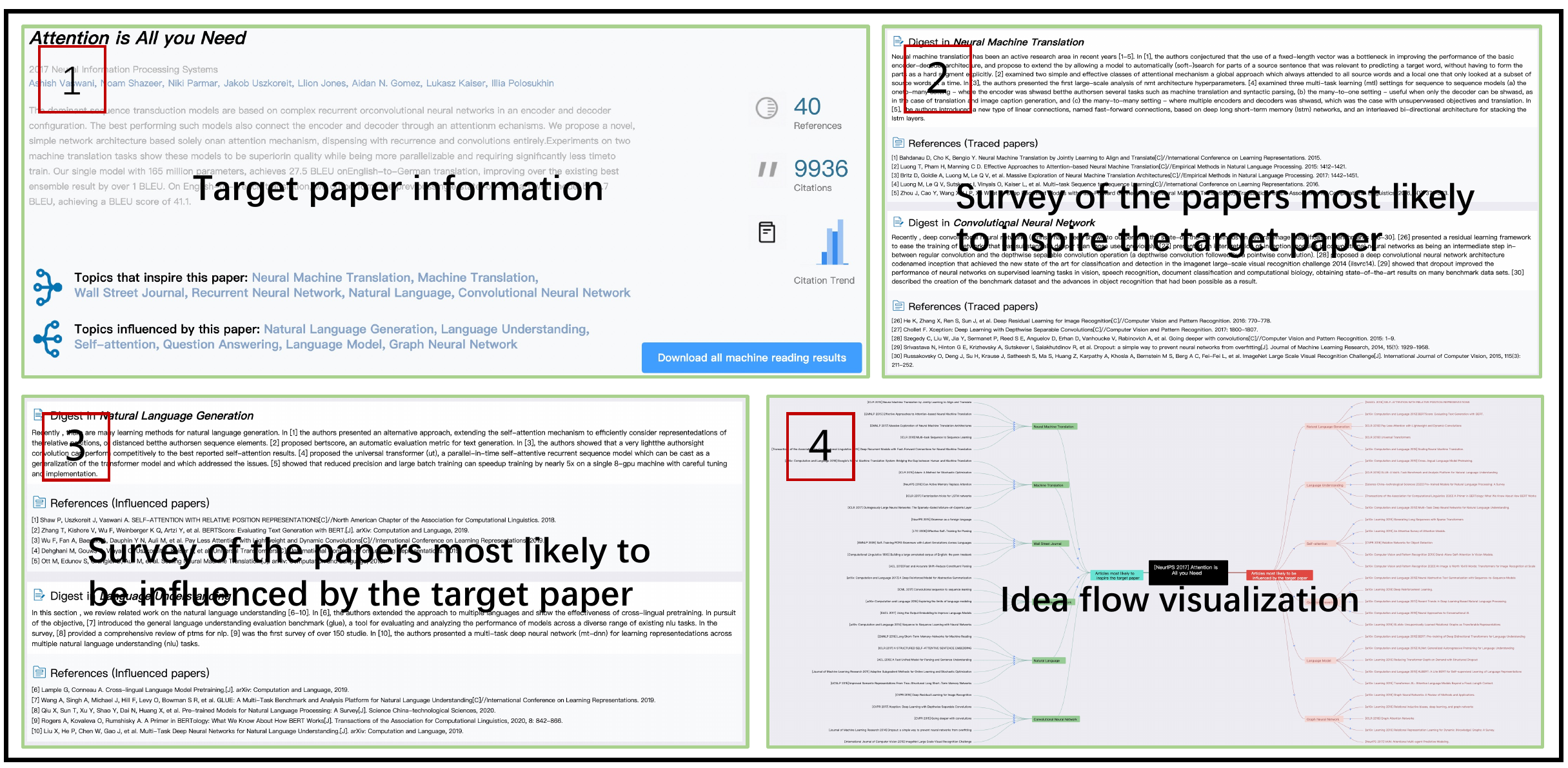} 
\caption{The front-end interface of the system.}
\label{fig2}
\end{figure}

We first introduce the algorithm pipeline (Fig. \ref{fig1}) of the IdeaReader, and then we introduce the front-end interface (Fig. \ref{fig2}) of our system.

\noindent \textbf{Paper query. }First, our system queries the candidate reference and citation papers of the target publication in the Acemap database \cite{tan2016acemap}. Specifically, taking reference papers as an example, IdeaReader first selects all first-order reference papers of the target publication. If the number of these papers does not exceed 100, it then selects all second-order reference papers (the papers cited by the first-order reference papers). Repeat this process until more than 100 papers are obtained. PageRank \cite{page1999pagerank} are used to obtain the top 100 papers as candidate papers among all the first-order or higher-order reference papers obtained. The selection of candidate citation papers follows a similar rule.

\noindent \textbf{Paper clustering and relevance scoring. }We then follow the approach in MRT \cite{yin2021mrt} to cluster and rank the candidate reference and citation papers separately. In this step, IdeaReader first combines TF-IDF with Sentence-BERT \cite{reimers2019sentence} to encode the abstracts of the papers and uses ProNE \cite{zhang2019prone} to incorporate the resulting embeddings according to the citation structure associated with the papers. After obtaining representations of papers, Kernel k-means \cite{kulis2009semi} is used to cluster candidate papers, and we treat each resulting cluster as a topic to inspire or be influenced by the target publication. To find out which papers are most likely to inspire or be influenced by the target publication under a specific topic, it utilizes the vector inner product and the actual citation relationship to calculate the relevance score between the candidate papers and the target publication. Finally, our system ranks the papers within each cluster according to their relevance scores.

\noindent \textbf{Survey generation under specific topics. }Finally, our system automatically summarizes the ideas of the top five papers with relevance scores in each cluster. The generated survey contains a heading, a general sentence, and summary sentences of the selected papers. IdeaReader uses automatic text annotation \cite{mei2007automatic} to extract the label from the title and abstract of the selected papers as the heading of the corresponding topic's survey. After comparison and human evaluation, we use BertSumABS \cite{liu2019text} to generate the general sentence from the abstracts of the selected papers. Our model is fine-tuned by the related work generation dataset provided by \citeauthor{chen2021capturing} (2021). Since we find that the results generated by BertSumABS are difficult to effectively summarize the input articles' ideas, we only keep the first sentence as the final general sentence. As for the summary sentences, we use SciBERT \cite{beltagy2019scibert} to train a binary classification model to extract these sentences from the abstracts of the input articles. Our dataset is constructed from the scientific abstract sentence classification dataset provided by \citeauthor{cohan2019pretrained} (2019) and \citeauthor{gonccalves2020deep} (2020). We utilize the sentences labeled as the `OBJECTIVE' class in the datasets as summary sentences, while the other sentences are used as negative samples to train the binary classification model. For readability, we finally align the subject and tense of the generated text.

\noindent \textbf{Front-end interface. }Our front-end interface is shown in Fig. \ref{fig2} and consists of four parts: (1) Target paper information panel. This part includes meta information of the target paper, references and citations statistics, and the topics that inspire or are influenced by the target publication. (2) Survey of the papers most likely to inspire the target paper. (3) Survey of the papers most likely to be influenced by the target paper. The panels in parts (2) and (3) contain surveys on multiple topics, each of which is presented in a single card. (4) Idea flow visualization. We call the result of visualization `Tracing and evolution tree'. In the tree, the root node is the target paper, the left branch is the topics and papers that inspire the target paper, the right branch is the topics and papers that are influenced by the target paper, and the direction of idea flow is from the left to right.

Our system supports the download of machine reading results as pdf files. The corresponding machine reading interface can be accessed via paper retrieval.

\section{Conclusion}
We presented IdeaReader, a machine reading system that helps researchers understand how scientific ideas flow between the target publication's contextual research. In the future, we will promote IdeaReader to scholars in different fields to get more comprehensive feedback for improving our system.

\end{document}